\documentclass[conference]{IEEEtran}
\IEEEoverridecommandlockouts

\usepackage{cite}
\usepackage{amsmath,amssymb,amsfonts}
\usepackage{algorithmic}
\usepackage{graphicx}
\usepackage{textcomp}
\usepackage{xcolor}
\usepackage{hyperref}

\usepackage{balance}
\usepackage{booktabs}
\usepackage{tabularx}
\usepackage{tikz}
\usepackage{subcaption}
\usepackage{newunicodechar}
\usepackage{listings}
\usepackage[most]{tcolorbox}

\definecolor{mycolor}{rgb}{0.98, 0.98, 0.98}
\definecolor{promptred}{rgb}{0.8, 0.0, 0.0}
\newcommand{\yeounoh}[1]{#1}
\newcommand{\shiva}[1]{#1}

\def\BibTeX{{\rm B\kern-.05em{\sc i\kern-.025em b}\kern-.08em
    T\kern-.1667em\lower.7ex\hbox{E}\kern-.125emX}}
\begin{document}

\title{High-Fidelity and Complex Test Data Generation for Google SQL Code Generation Services}

\author{\IEEEauthorblockN{Shivasankari Kannan}
\IEEEauthorblockA{
\textit{Google}\\
Mountain View, USA \\
shivasankari@google.com}
\and
\IEEEauthorblockN{Yeounoh Chung\IEEEauthorrefmark{1}}
\IEEEauthorblockA{
\textit{Google}\\
Mountain View, USA \\
yeounoh@google.com}
\and
\IEEEauthorblockN{Amita Gondi}
\IEEEauthorblockA{
\textit{Google}\\
Mountain View, USA \\
amitagondi@google.com}
\and
\IEEEauthorblockN{Tristan Swadell}
\IEEEauthorblockA{
\textit{Google}\\
Mountain View, USA \\
tswadell@google.com}
\and
\IEEEauthorblockN{Fatma \"{O}zcan}
\IEEEauthorblockA{
\textit{Google}\\
Mountain View, USA \\
fozcan@google.com}
\thanks{\IEEEauthorrefmark{1} Corresponding author.}
}

\maketitle

\begin{abstract}
The demand for high-fidelity test data is paramount in industrial settings where access to production data is largely restricted. Traditional data generation methods often fall short, struggling with low-fidelity and the ability to model complex data structures and semantic relationships that are critical for testing complex SQL code generation services like Natural Language to SQL (NL2SQL). In this paper, we address the critical need for generating syntactically correct and semantically relevant high-fidelity mock data for complex data structures that includes columns with nested structures that we frequently encounter in Google workloads.
We highlight the limitations of existing approaches used in production, particularly their inability to handle large and complex data structures, as well as the lack of semantically coherent test data that lead to limited test coverage. 
We demonstrate that by leveraging Large Language Models (LLMs) and incorporating strategic pre- and post-processing steps, we can generate syntactically correct and semantically relevant high-fidelity test data that adheres to complex structural constraints and maintains semantic integrity to the SQL  test targets (queries/functions). 
This approach supports comprehensive testing of complex SQL queries involving joins, aggregations, and even deeply nested subqueries, ensuring robust evaluation of SQL code generation services, like NL2SQL and SQL Code Assistant.
Our results demonstrate the practical utility of an LLM (\textit{Gemini}) based test data generation for industrial SQL code generation services where generating high-fidelity test data is essential due to the frequent unavailability and inaccessibility of production datasets for testing.
\end{abstract}


\section{Introduction}
In the fast-paced landscape of modern data systems, particularly within SQL development and testing, the demand for high-quality mock data that accurately mirrors real-world scenarios is more critical than ever \cite{whiting2008creating,gartner_marketguide,bhangdiya2015xda,chandra2015data,somwase2024data,arcuri2019restful}. This need is amplified in industrial settings where data is often confidential and/or sensitive. In such environments, direct access to production data is severely restricted, and traditional approaches like production data sampling become time-consuming, infeasible, or violate data privacy requirements. Historically, mock data generation has relied on manual creation, production data sampling or mathematical models \cite{somwase2024data,gartner_marketguide,9155857, soltana2017synthetic}. 

One key challenge for effective test data generation in industrial contexts is ensuring high fidelity.
High-fidelity data is syntactically correct and structurally compliant, accurately reflecting complex nested data structures and the nuanced relationships and correlations between fields and tables, as if it were sampled directly from production environments. 
This is crucial for supporting test cases with complex data structures that are consistent and satisfy data integrity requirements \cite{somwase2024data,chandra2015data,he2024verieql,bhangdiya2015xda}. 
For industrial applications heavily reliant on structured data  types such as JSON, XML, protocol buffers (a.k.a. proto or protobuf~\cite{protobuf}), it is paramount that the generated mock data strictly adheres to the specified structure and maintains data integrity and consistency \cite{endres2022synthetic,hernandez2022synthetic}. 
Specifically, our SQL code generation services utilize GoogleSQL~\cite{googlesql} that supports complex nested data types, like array, struct and proto -- making it a representative case for the challenges faced by many modern SQL dialects that support nested data structures.
Conventional test data generation methods, such as manual test data generation by the developers, are highly susceptible to errors that can compromise the validity of testing and development. Furthermore, scalability and efficiency of generation are critical for enterprise needs, especially when a large amount of test data is needed \cite{gretel_definitiveguide,mockaroo,hackolade_generatemockdata}. As enterprise software systems grow in scale and complexity, the need for large volumes of mock data increases exponentially. Manual and heuristic data generation approaches become impractical and unsustainable in such environments, hindering development velocity and quality assurance efforts.

The generation also needs to ensure data value diversity to effectively simulate the wide range of scenarios encountered in real-world systems, including both common use cases and rare edge cases.
Existing data generation methods often fall short in capturing this diversity, while rule-based systems tend to be inflexible and limited in their ability to generate varied data that reflects the complexities of industrial data distributions \cite{mannino2019real,hackolade_generatemockdata,patki2016synthetic,mockaroo}.

The emergence of Large Language Models (LLMs) presents a paradigm shift, offering new avenues for automating the generation of mock data, especially for structured formats such as database schema and protocol buffers. 
For clarity, throughout this paper, we use the term "schema" to generally refer to a formal structural definition, encompassing both Google SQL schema definitions and/or Protocol Buffer (proto) definitions with potentially nested fields.
LLMs can generate mock data that more closely approximates real-world data distributions and patterns based on their acquired real-world knowledge. LLMs can capture subtle nuances and semantic relationships between data attributes that are difficult to replicate with traditional methods \cite{xu2024llms,sui2024table,borisov2022language}.
This is crucial for comprehensive production testing while also avoiding the privacy risks associated with using actual production data (regulatory security and privacy policy restricts access to user data). Additionally, LLMs can understand SQL structures and facilitate structurally compliant and schema-aware data generation \cite{singh2024exploring,borisov2022language,sui2024table,nguyen2024generating}. 

While the use of LLMs for synthetic data generation is becoming popular in various domains \cite{singh2024exploring,gretel_definitiveguide,lu2023machine,karmarkar2024navigating,liu2023fill,zhu2025bare,li2023synthetic,patel2024datadreamer}, the generation of structured data introduces unique challenges that are particularly relevant to industrial applications. 
A significant limitation of current LLM-based data generation approaches lies in their ability to effectively handle more complex data structures commonly found in our workloads. Specifically, LLMs struggle to accurately generate data that conforms to large (many columns or fields) and complex nested structures (Table \ref{tab:customer_table_data_generation}). While LLMs can generate data that adheres to basic and flat schema structures, accurately generating data that conforms to deeply nested structures (e.g., protobuf, GoogleSQL) with complex dependencies remains a challenge. This limitation is particularly problematic in testing real-world SQL generation services that rely on intricate data models, where the nuances of nested relationships over very wide tables are critical for comprehensive testing and validation in production.

A further crucial challenge is ensuring the semantic validity of the generated data, especially when leveraging Large Language Models (LLMs). While LLMs might generate data that is syntactically and structurally correct, it often lacks the necessary semantic coherence across columns and tables. This occurs because many existing test data generation approaches focus primarily on approximating per-column or pairwise column distributions from production data \cite{zhang2023mixed,lee2023codi,liu2023fill,mannino2019real,whiting2008creating,bhangdiya2015xda,chandra2015data,somwase2024data}. Whether using statistical or vanilla LLM methods \cite{sui2024table,nguyen2024generating}, this limited focus frequently leads to illogical combinations of column values, semantically incoherent join tables, and data inconsistencies—all of which violate real-world business rules.

Given a schema or proto definition, LLMs can be guided to generate mock data that adheres to the specified structure and constraints. However, tabular data generation via language models remains an active and evolving field, facing challenges such as accurately capturing complex inter-column (pairwise) relationships, mitigating the risk of ``hallucinations'' or data inconsistencies, and effectively handling rare values and outliers (also see Table~\ref{tab:customer_table_data_generation} for vanilla LLM evaluation results)~\cite{sui2024table,nguyen2024generating}.
Overcoming these limitations, including both structural and semantic challenges, is crucial for fully realizing the potential of LLMs in generating comprehensive and accurate mock data for complex workloads.

In this paper, we address the critical need for generating syntactically correct and semantically relevant high-fidelity mock data for structurally well-defined, yet complex data formats, focusing on the specific requirements of industrial use cases. We define high-fidelity data as that which is syntactically accurate and also semantically aligned with the testing context, ensuring the generated data adheres to the constraints and logic implied by the test target. 
A key advantage is that, given a protocol buffer (proto) definition, LLMs can be guided to generate mock data that adheres to the specified structure and constraints. This includes respecting all implicit and explicit constraints, such as generating data within a specific date range when the SQL query under test includes a corresponding filter (e.g., WHERE year = 2022). This high-fidelity data is essential to support a wide range of complex test cases for services like NL2SQL and SQL code assistance systems. Robust evaluation of such generation systems requires this high-fidelity and complex test data, which is often inaccessible due to privacy concerns or a lack of sophisticated generation capabilities. Services like NL2SQL support arbitrarily complex use cases, generating intricate SQL queries involving joins, aggregations, and nested subqueries, for which vanilla out-of-the-box LLMs simply fail to generate suitable, high-quality test data.

Our key contributions are:
\begin{itemize}
    \item \textbf{LLM-based High-Fidelity and Efficient Test Data Generation:} We present a novel approach leveraging LLMs, enhanced with pre- and post-processing, to generate high-fidelity test data compliant with large complex structures, like production protobuf and GoogleSQL schema that include nested fields. Our generation method is significantly more efficient than manual generation used in production where production data access is restricted.
    \item \textbf{Enhanced Test Coverage for Complex SQL Queries:} We demonstrate that our generated test data significantly improves test coverage of Google SQL code generation services, where many test cases are disabled due to limited access to production data and the difficulties of manual data generation. Our fundamental approach is SQL dialect-agnostic and should be readily applicable to other SQL code generation services that face similar challenges.
    \item \textbf{Semantic Coherence With Test SQL Queries:} We emphasize and achieve the generation of semantically relevant test data. We provide a comparative analysis against production data sampling, showcasing the ability of our generated test data to cover edge cases and include semantically relevant predicate columns.
\end{itemize}

The remainder of this paper is structured as follows. Section 2 discusses real-world SQL code generation services that serve as use cases for our work. Section 3 presents our proposed method for generating high-quality mock data. Section 4 describes our experimental setup and presents the evaluation results. Section 5 discusses related work in the field of data generation. Section 6 provides a discussion of our findings and their implications. Finally, Section 7 concludes the paper.

\section{SQL code generation services}
\label{sec:sql_code_generation_services}

In this section, we examine the challenges of generating high-fidelity test data for two real-world GoogleSQL \cite{googlesql} SQL code generation services: SQL Code Assistant and NL2SQL. 
\yeounoh{A critical requirement for testing these services is the ability to handle the rigid structural constraints of GoogleSQL, which supports deeply nested types like array, struct, and proto. Traditional constraint-based generators~\cite{mannino2019real,hackolade_generatemockdata,patki2016synthetic,mockaroo}, often built for flat, relational schemas, cannot natively parse these complex nested structures. Because these tools lack the logic to maintain semantic integrity across multiple levels of nesting, they are unable to produce the syntactically valid instances required for executing the SQL functions under test.}

\begin{table}[t!]
\centering
\begin{minipage}{\columnwidth}
\caption{Traditional test data generation methods for real-world SQL code generation services. Test coverage is largely limited by the test data quality, especially when the underlying data structure is large and complex with deep nesting.}
\label{tab:sql_services_comparison}
\resizebox{\columnwidth}{!}{%
\begin{tabular}{lccc}
\toprule
 & SQL Code Assistant & NL2SQL \\
\midrule
Schema complexity & High & High \\
Test (SQL query) complexity & Low & High \\
Data generation method & Manual & Sampling\footnote{restricted to a limited set of domains/tables in production}  \\
\midrule
Test coverage & Low & Low \\
\bottomrule
\end{tabular}%
}
\end{minipage}%
\end{table}

Table \ref{tab:sql_services_comparison} summarizes test \& schema complexities, as well as the traditional test data generation methods used for the services. The underlying data schema for both services are complex with many columns/fields with nested fields (nested proto and array types). The test complexity for NL2SQL is even more complex as the service generates SQL queries spanning over multiple tables across many customer domains. Test cases are  submitted by the internal developers and also by crowdsourcing. The test SQL queries for SQL Code Assistant are mostly simple (single-table) selection queries; however, the manually curated test data overlooks complex nested structures from the schema limiting the test cases to SQL queries targeting mostly flat schemas with non-nested fields. 
The SQL query complexity vary, but most of them use nested fields and subqueries with joins. SQL code generation services require high-fidelity test data that adheres to the complex schema  with nested structures that support complex and diverse sets of SQL queries/functions. Additionally, NL2SQL service evaluation requires testing if predicted SQL generates a result set that is semantically coherent with the natural language question. This also means that the test data instances should be semantically relevant and aligned with the context of the posed data question. Unfortunately, production sampling, though anonymized, is restricted to a few accessible domains and tables, resulting in incomplete test scenarios and low test coverage.

\begin{table*}[t!]
\centering
\caption{Evaluation of test data generation for a large Google Ads production customer table with 72 columns (65 are nested). With production data restricted, manual and vanilla LLM methods fail to fully address this moderately complex schema. Our method achieved near-perfect results across all criteria.}
\label{tab:customer_table_data_generation}
\begin{minipage}{\textwidth}
\begin{tabularx}{\textwidth}{lXXX}
\toprule
& \multicolumn{3}{c}{Test Data Generation Methods}\\ 
\cmidrule(r){2-4}
Evaluation Criteria & Manual & Vanilla LLM & Our Method \\
\midrule
Primary key generated & Yes & No & Yes \\
Number of fields with correct field names & 11/72 & 62/72 & 72/72 \\
Total number of table columns generated & 11/72 & 71/72 & 72/72 \\
\midrule
Number of nested fields with incorrect level of nesting generated & 0/10 & 36/64 & 0/64 \\
Number of nested fields with at least one correct field name & 10/10 & 13/64 & 64/64 \\
Number of nested fields with correct enum values generated & 10/10 & 2/50 & 50/50 \\
Number of nested fields with correct scalar values generated & 10/10 & 5/64 & 64/64 \\
Number of nested fields with all the column values generated & 9/10 & 0/64 & 64/64 \\
\midrule
Generation time per instance (50p) & 1 hour\footnote{A domain-expert engineer required over one hour to manually curate a single, comprehensive test instance. Given the schema complexity, manual generation is practically infeasible for large-scale industrial testing.} & 37 sec & 167 sec \\
\bottomrule
\vspace{-5mm}
\end{tabularx}
\end{minipage}
\end{table*}

\subsection{SQL Code Assistant}

SQL Code Assistant is designed to aid developers in writing SQL queries or SQL functions that access data models (schema, proto). Its testing involves the generated SQL queries/functions and test data instances that are often manually prepared and verified. This manual data generation often leads to incomplete data instances with partially filled schema elements neglecting complex and nested structures. This can lead to low test coverage, limiting the range of executable test queries. Furthermore, even a simple test case over a primitive (char) column can have low coverage if the data instances do not cover all the edge/branch cases under test.

\begin{figure}[!htbp]
\centering
\begin{tikzpicture}
   \node[draw=black!60, fill=lightgray!5, rectangle, rounded corners, inner sep=6pt, text width=0.92\columnwidth] {
            \begin{lstlisting}[basicstyle=\ttfamily\scriptsize, columns=flexible, showstringspaces=false]
CREATE PUBLIC FUNCTION GetBalanceInUsd(Input fake_table) RETURNS INT64
AS (
  CASE
  WHEN private_info.running_balance.currency = 'USD' THEN private_info.running_balance.amount
  WHEN private_info.running_balance.currency = 'GBP' THEN private_info.running_balance.amount * 1.26
  WHEN private_info.running_balance.currency = 'EUR' THEN private_info.running_balance.amount * 1.05
  ELSE NULL
  END
);
            \end{lstlisting}
        };
\end{tikzpicture}
\caption{Simple selection query accessing deeply nested fields on a fake table, \textit{currency} and \textit{amount}.}
\label{fig:simple_selection_query}
\end{figure}

Figure~\ref{fig:simple_selection_query} illustrates a simple selection query function accessing nested columns (nested proto type fields, \textit{fake\_table} $\rightarrow$ \textit{private\_info} $\rightarrow$ \textit{running\_valance}).
The table schema is very large with 161 columns and the column (proto type) that this query is accessing has 3 levels of nesting. The query function is tested using manually curated test data instances that are often partially filled with a few fixed test values on the required fields and/or default values on the nested structures and fields. We found that this leads to low test coverage on many user defined SQL functions for the SQL Code Assistant service. For instance, the above example requires at least three test data instances with all three distinct \textit{currency} column values, ``USD", ``GBP" and ``EUR" for the optimal test coverage. As the test complexity increases with more predicates and nesting, providing all test data instances with all the relevant test values becomes much harder. As a result, this was limiting the complexity of test cases submitted by the developers who are responsible for both test writing and data generation. Most of the test queries are simple single-table selection or aggregation queries with a handful of predicates.

\subsection{NL2SQL}
NL2SQL is a more advanced service that translates natural language queries into complex SQL statements. It supports highly complex table schema with foreign key relationships; the generated SQL queries involving joins, aggregations, and nested subqueries over multiple tables. NL2SQL testing also requires the test data to be semantically coherent with the natural language question and the SQL. 
Due to privacy constraints and the scale of production data, NL2SQL testing relies on sampling from a restricted set of domains and tables. This limited production sampling significantly hampers test coverage and the ability to thoroughly evaluate across diverse scenarios.
More specifically, the service has accumulated a large set of test SQL queries that are submitted by the internal developers, crowdsourcing and also provided by the customers with golden queries.

\begin{figure}[!htbp]
\centering

\begin{subfigure}{\columnwidth}
    \centering
    \textbf{\small (a) Natural Language Question} \\
    \vspace{1mm}
    \begin{tikzpicture}
        \node[draw=black!60, fill=lightgray!5, rectangle, rounded corners, inner sep=6pt, text width=0.92\columnwidth] {
            \small \textit{"Which fake\_devices have tasks related to fake\_task\_name with a start date after 2020-06-01, and what are the tasks and their start dates?"}
        };
    \end{tikzpicture}
\end{subfigure}

\vspace{3mm}

\begin{subfigure}{\columnwidth}
    \centering
    \textbf{\small (b) Golden GoogleSQL Query} \\
    \vspace{1mm}
    \begin{tikzpicture}
        \node[draw=blue!40, fill=blue!2, rectangle, rounded corners, inner sep=6pt, text width=0.92\columnwidth] {
            \begin{lstlisting}[basicstyle=\ttfamily\scriptsize, columns=flexible, showstringspaces=false]
SELECT name, ARRAY_AGG(STRUCT(task, start_date))
FROM (
  SELECT d.name, t.name AS task,
    FORMAT_TIMESTAMP("%Y-%m-%d", 
      PARSE_TIMESTAMP("%Y-%m-%dT%H:%M:%E*S%Ez", t.start_date)) AS start_date
  FROM fake_project_tasks AS t
  INNER JOIN fake_devices AS d ON t.project_id = d.id
  WHERE t.name LIKE '%fake_task_name' AND t.start_date > "2020-06-01"
  ORDER BY start_date
) GROUP BY name;
            \end{lstlisting}
        };
    \end{tikzpicture}
\end{subfigure}

\caption{NL2SQL test case illustrating (a) a natural language question and (b) its corresponding golden SQL query involving a join within a subquery and complex timestamp formatting.}
\label{fig:nl2sql_nested_query}
\end{figure}

Figure~\ref{fig:nl2sql_nested_query} illustrates a complex test case that require evaluating a golden SQL query with a join in the subquery. Some of the columns in the schema are also nested (proto and array types). Production sampling (graph mining-based) tends to focus on getting the most diverse set of instances, which may not guarantee a test data sample that is relevant for the given question. Given relatively small sample sizes, many of the test queries result in result sets with fewer rows than expected -- yielding unrepresentative evaluation results. In our case, we generate test data that adheres to the arbitrarily complex schema structures and also relevant to the given test SQL structures using LLM. It is important to note that our generation process takes into account the semantics of the question and the schema. For instance, the LLM can make connection between the fake project tasks in the question and the fake device  related task's start date (\textit{t.start\_date}) to build up the context for relevant data generation. We observe that test data generated with the right semantic context can actually improve the test coverage significantly compared to production samples. 

\subsection{\yeounoh{Testing Requirements and Evaluation Methodology}}
\yeounoh{Industrial SQL code generation services require more than just syntactically valid test data; they demand a high degree of semantic fidelity to ensure evaluation validity. The primary evaluation metric for these systems is Execution Accuracy (EX), which assesses the model's performance by executing both the predicted SQL and the golden SQL on a database and comparing their output result sets~\cite{chung2025long,floratou2024nl2sql}. 
For this evaluation to be meaningful, the underlying test data (generator) must satisfy the following critical requirements. 
First, due to strict security policies, the generator cannot have access to production data samples. Instead, it must build the right context by synthesizing metadata, including source code comments, schema definitions, documentation, and the test SQL queries/questions themselves.
Second, the data must be generated with values that satisfy the predicates in the SQL queries, so that the result sets are not empty. Semantically irrelevant data can lead to empty result sets for both predicted and golden queries, producing trivial matches that undermine the evaluation's validity.
}

\subsection{Problem Definition}
Let $\Lambda$ be a schema definition, like table schema or proto, with various types of fields/columns $\Xi$. The column set $\Xi$ includes primitive type (e.g., numerical, string, boolean) $\xi^{prim}$, categorical type $\xi^{cat}$ (e.g., enum), and nested type (e.g., struct, array, proto) $\xi^{nest}$ . 
$\xi^{nest}$ can be another (nested) schema definition $\Lambda_i$ that can have more nested structures within $\xi^{nest}_i$.
Our work focuses on generating arbitrarily complex test data with any number of columns and with deeply nested structures, $\Xi = \{\xi^{prim}_1, \xi^{cat}_2, \xi^{nest}_3, ..., \xi_n\}$, for testing test SQL queries $\Omega = \{\omega_1, \omega_2, ..., \omega_m\}$ and also without any training tabular dataset $\tau_{\Lambda}$ that confirms to $\Lambda$ -- due to restricted production data access in real-world SQL code generation services.

Specifically, we leverage a pre-trained LLM, \textit{Gemini}, to generate high-fidelity and complex $\tau_{\Lambda}$ that enables testing $\Omega$.  Table~\ref{tab:customer_table_data_generation} illustrates the evaluation of test data generation for a large complex schema. Using the out-of-the-box \textit{Gemini} model (\textit{Vanilla LLM}) for mock test data generation with $\Lambda$ resulted in incomplete data. Our generation method, incorporating strategic pre- and post-processing steps, results in the most complete test data, better than the manually prepared one by a human expert in an hour.  Production sampling is not used, since the table is restricted. While the traditional manual data generation is very time-consuming, the LLM based generation runs much faster. 
We further discuss the per-query test data generation time and cost in Section~\ref{sec:cost_analysis}.

\section{Test data Generation for SQL code generation services}

\begin{figure}[!t]
\centering 
\includegraphics[width=\columnwidth]{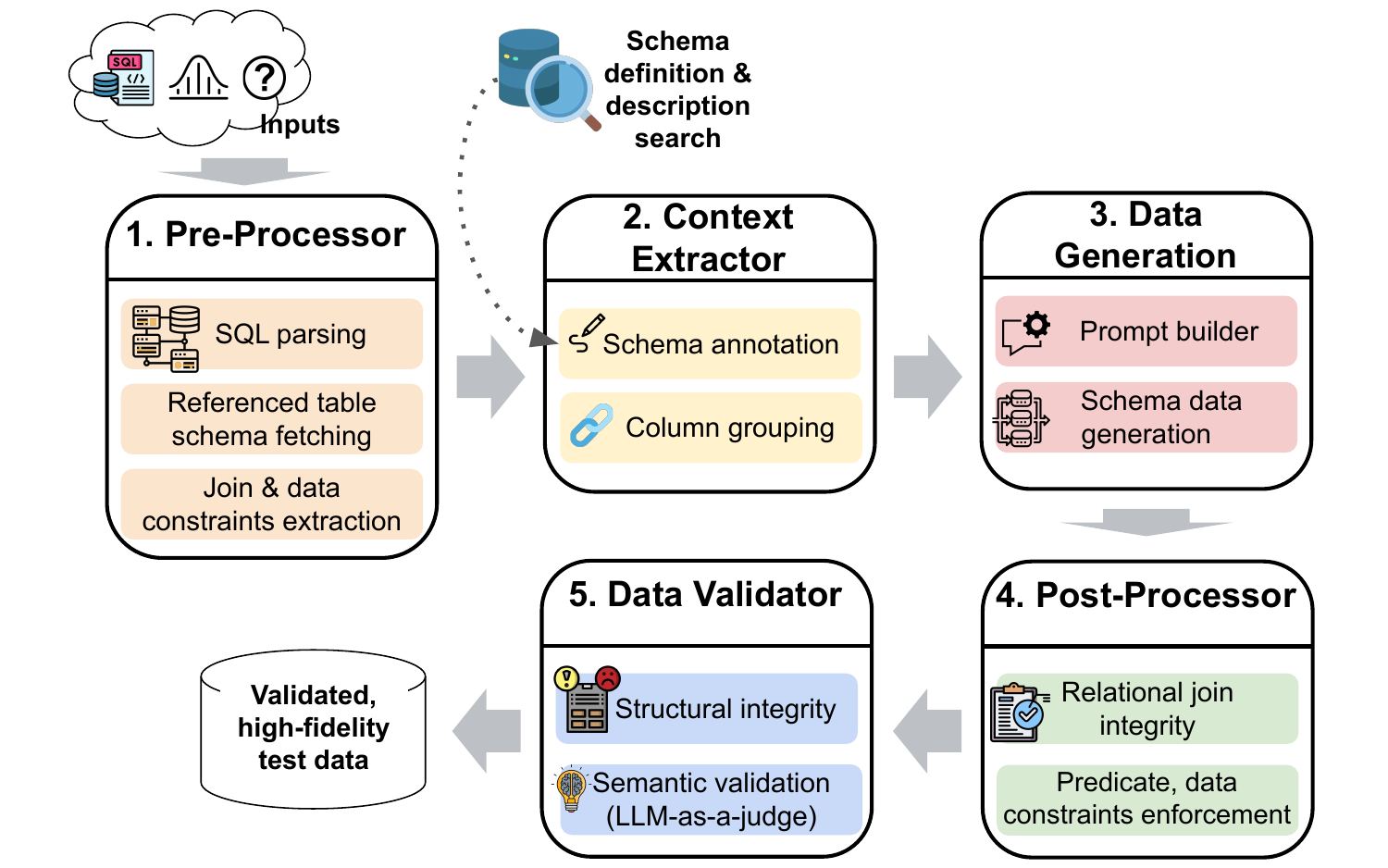} 
\caption{\yeounoh{LLM (Gemini)-based test data generation process: User provides context map (questions, schema, additional criteria) and test body (SQL query) as input. This data generation request triggers a series of data processing steps and modules. The goal is to generate syntactically correct and semantically relevant test data for SQL code generation services, like NL2SQL. We leverage the latest LLM capabilities to do this efficiently at scale.}}
\label{fig:system_overview}
\end{figure}

To automate the generation of test data that aligns with user-specified criteria, we propose a novel test data generation pipeline leveraging the power of LLMs (\textit{Gemini}). Figure~\ref{fig:system_overview} provides a high-level overview of this process, illustrating the flow of data and processing from user input to the final validated mock test data output. The process begins with the user providing a context map -- containing questions, schema definitions, and additional criteria—along with a test body, such as an SQL query or SQL function implementation. This input triggers a series of modules within the data generation library, including a pre-processor, context extractor, prompt builder, post-processor, and data validator. Each module plays a crucial role in refining and validating the generated data, ensuring it meets both the realized criteria from the schema and the query, as well as the user provided criteria.

\subsection{Pre-processor}
This module will seed the signals needed to generate high-fidelity data, automatically based on information from the context map and the test SQL query body. Additionally this can also be passed from the user. Any user provided input will override auto-generated ones. The signal will be a map of column name to column value (literal value, range or values, or a generator like incremental or other distribution, e.g., use a uniform distribution from 0 to 1 to populate this numeric column).
These signals would be added to the context map and used later for post-processing.
\\
\noindent\textbf{Signals for Data Generation: }
Additionally the following signals would be needed to generate high-fidelity data:
\begin{itemize}
\item Number of rows required.
\item Constraints in data points - e.g. specifying a currency in the example provided.
\item Column correlations.
\item Variations in data points - e.g. any field that needs to be different in each data point like customer ID.
\item SQL query and question (if applicable)
\end{itemize}
\vspace{1mm}

\begin{figure}[!htbp]
\centering

\begin{subfigure}{\columnwidth}
    \centering
    \textbf{\small (a) Input SQL Query} \\
    \vspace{1mm}
    \begin{tikzpicture}
        \node[draw=black!60, fill=lightgray!5, rectangle, rounded corners, inner sep=6pt, text width=0.92\columnwidth] {
            \begin{lstlisting}[basicstyle=\ttfamily\scriptsize, columns=flexible, showstringspaces=false]
SELECT * FROM (
  SELECT * FROM fake_table_1
  WHERE fake_column = "Regional_Team_Americas"
  AND date >= "2023-01-01" AND date <= "2023-03-31"
) AP
LEFT JOIN (SELECT * FROM fake_table_2) AS Capacity
ON SAFE_CAST(AP.date AS STRING) = Capacity.week_start_date 
AND AP.username = Capacity.username
            \end{lstlisting}
        };
    \end{tikzpicture}
\end{subfigure}

\vspace{3mm}

\begin{subfigure}{\columnwidth}
    \centering
    \textbf{\small (b) Extracted Generation Targets} \\
    \vspace{1mm}
    \begin{tikzpicture}
        \node[draw=blue!40, fill=blue!2, rectangle, rounded corners, inner sep=6pt, text width=0.92\columnwidth] {
            \begin{lstlisting}[basicstyle=\ttfamily\scriptsize, columns=flexible, showstringspaces=false]
T1: ["fake_table_1"] 
    Constraints: "fake_column='Regional Team Americas' 
    AND date>='2023-01-01' AND date<='2023-03-31'"
T2: ["fake_table_2"]
    Constraints: []
            \end{lstlisting}
        };
    \end{tikzpicture}
\end{subfigure}

\vspace{3mm}

\begin{subfigure}{\columnwidth}
    \centering
    \textbf{\small (c) Extracted Join Constraints} \\
    \vspace{1mm}
    \begin{tikzpicture}
        \node[draw=green!60, fill=green!2, rectangle, rounded corners, inner sep=6pt, text width=0.92\columnwidth] {
            \begin{lstlisting}[basicstyle=\ttfamily\scriptsize, columns=flexible, showstringspaces=false]
[{"fake_table_1": "date", "fake_table_2": "week_start_date"},
 {"fake_table_1": "username", "fake_table_2": "username"}]
            \end{lstlisting}
        };
    \end{tikzpicture}
\end{subfigure}

\caption{\yeounoh{The Pre-processor parses the input SQL query (a) to extract semantic seeds for the data generation pipeline. This process identifies individual table references and their associated filtering predicates (b), which serve as hard constraints for generating independent table rows. Furthermore, join conditions are extracted (c) to define the necessary column associations between tables, ensuring relational consistency across the final generated data.}}
\label{fig:preprocessor_extraction}
\end{figure}

\yeounoh{
\noindent\textbf{Handling Subqueries and Table-Level Parallelism:}
Our framework decomposes complex SQL queries into atomic data generation targets based on the underlying table schemas. This allows for the parallelization of the generation process across independent tables referenced in the main query and its nested subqueries.
To maintain semantic integrity across the entire query tree, the pre-processor extracts data generation signals (e.g., predicate constraints) and maps them to each referenced table. Independent tables are flagged as primary generation targets under \textit{``Subqueries:"} in the pre-processor output, while redundant or dependent references are deduplicated to optimize token usage and generation time (see  Figure~\ref{fig:preprocessor_extraction} for an illustration). For multi-table queries involving joins, the pre-processor isolates join-key associations, which are subsequently enforced during the deterministic post-processing phase to ensure relational consistency.}
\\
\noindent\textbf{Handling Joins:}
Joins information is extracted from the SQL query separately to show table and column association for the join. The example SQL above the join information will be extracted as in Figure~\ref{fig:preprocessor_extraction}. Our method generates data (rows) per schema given a SQL query, the extracted join information will then be used in the post-processing to match the values in the table and columns present in the Join Constraints data structure. It is not possible to capture these nuances in a naive approach using Vanilla LLM.

\subsection{Context extractor}
The context extractor module generates context relevant to the schema. 
\yeounoh{
Industrial schemas often suffer from cryptic (non-human readable) identifiers and a lack of formal CHECK constraints, creating a significant challenge for traditional rule-based generators.
To bridge this gap, we use a separate system to aggregate human-written metadata, including schema descriptions, source code comments, and technical documentation.
These comments are available at the schema/proto level and also individual column level. 
Because this information is often fragmented and under-specified, we use the LLM to make these comments consistent and fill in missing column descriptions~\cite{narayan2022can,korini2023column,li2024automatic,peeters2023schema}. 
This interpolation task is essential for our zero-shot setting (i.e., no production data seeding to learn the distributions).
While we acknowledge that LLM-generated annotations are not perfect, they provide the necessary semantic grounding to generate executable and coherent test data where other tools fail.
}

Beyond metadata, the module extracts functional signals directly from the test SQL query, as illustrated in Figure~\ref{fig:preprocessor_extraction}, so that the model can explore and generate different values a relevant column in the predicates should take for the test (for instance, the LLM understands that it should generate for \textit{currency} $\in \{``USD", ``GBP", ``EUR"\}$ from the query in Figure~\ref{fig:simple_selection_query}). The same set of constraints are later used for the semantic quality verification of the generated data as well. In Section~\ref{sec:business_analytics_discussion} we discuss another example where our LLM-based generation successfully generate test data around $date\_trunc(logdate, quarter)$ constraint, leaving no quarter without test data.

\subsection{Prompt builder}

\begin{figure}[!th]
\centering
\caption{Data generation prompt template. Custom generation seeding instructions can be passed as \textit{data\_generation\_signals} and \textit{user\_instruction}.}
\label{fig:prompt_template}
\begin{tikzpicture}
    \node[draw=black!60, fill=lightgray!5, rectangle, rounded corners, inner sep=6pt, text width=0.92\columnwidth] {
        \begin{minipage}{\columnwidth}
            \scriptsize 
            \textbf{You are a GoogleSQL expert.} GoogleSQL is a Google internal implementation of the SQL query language.
            Given a proto description, generate sample data in textproto format.
            \\[0.5em]
            Use only the proto names present in the proto files.
            Do not create columns that do not exist. Check if the columns match with the proto description.
            \\[0.5em]
            \textbf{Instructions for generating sample data:}
            \begin{itemize}
                \item Pay careful attention to the datatypes in the proto columns.
                \item Don't make up the proto field names.
                \item Carefully match the proto field names to their data types.
                \item Don't repeat the fields more than once.
                \item Ignore any deprecated fields in the proto.
                \item Generate the nested proto fields in the proto file.
            \end{itemize}

            \textbf{Instructions for specific columns in the given schema:}
            \{constraints\}
            \\[0.5em]
            \textbf{Here are the steps that each question should follow:}
            \begin{enumerate}
                \item Generate the proto column name only if present in the proto file.
                \item If you have already generated data for a field, please move to the next field.
                \item Generate non-zero values for every numeric field.
            \end{enumerate}
            
            \vspace{0.5em}
            
            \begin{lstlisting}[
             basicstyle=\color{promptred}\ttfamily\scriptsize,
             columns=fullflexible,
             breaklines=true,
             breakatwhitespace=true,
             frame=none,
             language=tex,
            ]
```
system_instruction=f"""
{system_prompt_prefix}
{data_generation_signals}
"""

user_instruction=f"""
Task: Given the following proto description, generate  {number_of_data_points} rows of sample data for {col_names} columns. {user_input}

Proto/schema Description:
{proto_description}
"""

mock_data_gen_prompt=f"""
system:{system_instruction}
user:{user_instruction}
model:
```
\end{lstlisting}
\end{minipage}
};
\end{tikzpicture}
\end{figure}

The test data generation prompt is constructed using the following template in Figure~\ref{fig:prompt_template}. The template contains some system instructions to specify the task and also to prevent common hallucination problems, like repeating the same column more than once, carefully validating and matching the column/field names to the schema. While structural hallucination (e.g., incorrect column/field names) are easier to detect and address almost perfectly in the subsequent post-processing and data validation steps, semantic hallucination (e.g., generating test data in fiscal year 2023, while the NL2SQL test question is about fiscal year 2025) is much harder to address completely, as further discussed in Section~\ref{sec:data_validator}.

\begin{figure}[!htbp]
\centering

\begin{subfigure}{\columnwidth}
    \centering
    \textbf{\small (a) Without Column Grouping} \\
    \vspace{1mm}
    \begin{tikzpicture}
        \node[draw=red!50, fill=red!2, rectangle, rounded corners, inner sep=6pt, text width=0.92\columnwidth] {
            \begin{lstlisting}[basicstyle=\ttfamily\scriptsize, columns=flexible, showstringspaces=false]
{
  "contract_created_date": "2023-08-15",
  "contract_effective_date": "2023-04-01", // ERROR
  "contract_closed_date": "2023-03-30",    // ERROR
  "contract_termination_date": "2023-12-31"
}
            \end{lstlisting}
        };
    \end{tikzpicture}
\end{subfigure}

\vspace{4mm}

\begin{subfigure}{\columnwidth}
    \centering
    \textbf{\small (b) With Column Grouping} \\
    \vspace{1mm}
    \begin{tikzpicture}
        \node[draw=green!60, fill=green!2, rectangle, rounded corners, inner sep=6pt, text width=0.92\columnwidth] {
            \begin{lstlisting}[basicstyle=\ttfamily\scriptsize, columns=flexible, showstringspaces=false]
{
  "contract_created_date": "2023-10-27",
  "contract_effective_date": "2023-11-01", // CORRECT
  "contract_closed_date": "2024-10-31",    // CORRECT
  "contract_months_from_effective_date": 12,
  "contract_termination_date": "2024-05-15"
}
            \end{lstlisting}
        };
    \end{tikzpicture}
\end{subfigure}

\caption{\yeounoh{Column grouping based on column correlation helps generating more coherent set of values. (a)  Without column grouping the contract effective start date is before the contract created date; (b) with column grouping  all dates are aligned and correct.}}
\label{fig:column_grouping}
\end{figure}

\begin{figure}[!htbp]
\label{fig:column_annoation}
\centering
\begin{subfigure}{\columnwidth}
    \centering
    \textbf{\footnotesize (a) Source Context (Human Comments \& Docs)}
    \begin{tikzpicture}
        \node[draw=black!60, fill=lightgray!5, rectangle, rounded corners, inner sep=5pt, text width=0.92\columnwidth] {
            \texttt{\scriptsize "terminated\_by\_clm\_system" // "If the contract was terminated, this field indicates the Contract Lifecycle Management System... part of the source information about this contract's lineage."}
        };
    \end{tikzpicture}
\end{subfigure}

\vspace{0.3cm}

\begin{subfigure}{0.48\columnwidth}
    \centering
    \textbf{\footnotesize (b) Without Annotation}
    \begin{tikzpicture}
        \node[draw=red!50, fill=red!2, rectangle, thick, text width=\linewidth, minimum height=1.2cm] {
            \texttt{\scriptsize "terminated\_by\_clm\_system": \\ \textbf{"TerminationSystem"}}
        };
    \end{tikzpicture}
    \textit{\tiny Generic/Hallucinated}
\end{subfigure}
\hfill
\begin{subfigure}{0.48\columnwidth}
    \centering
    \textbf{\footnotesize (c) With Annotation}
    \begin{tikzpicture}
        \node[draw=green!60, fill=green!2, rectangle, thick, text width=\linewidth, minimum height=1.2cm] {
            \texttt{\scriptsize "terminated\_by\_clm\_system": \\ \textbf{"InternalCLMSystemCode"}}
        };
    \end{tikzpicture}
    \textit{\tiny Semantically Correct}
\end{subfigure}

\caption{\yeounoh{The context extractor fetches (a) human-written documentation and source code column annotations to guide the LLM. (b) Without this, the LLM generates generic strings; (c) with annotation, it produces valid values.}}
\label{fig:column_annoation}
\end{figure}

When we make the LLM call for generation, we make the request for each nested field separately on a separate thread (see Section~\ref{sec:cost_analysis} for more details). For the primitive/scalar fields, we group and pass one generation call to handle 30 fields at a time.\shiva{ The grouping is based on the column correlations identified by the context extractor. This helps maintain semantic consistency across the correlated columns, as supposed to generating a valid value for each column individually (see Figure~\ref{fig:column_grouping}). 
}
The column grouping sometimes leads to unexpected hallucinations where the LLM introduces fake (non-existent) columns, especially when the schema is small with a fewer number of primitive fields (see Section~\ref{sec:error_analysis} for an example). To guard against such hallucination, we also keep track of the requested column generations so that any hallucinated columns can be detected and removed.
\shiva{The schema (proto description) passed here also contains column/field annotations prepared by the context extractor. The annotations are attached to the corresponding schema elements as comments. This also improves the semantics of the generated data, as illustrated in Figure~\ref{fig:column_annoation}.}

\subsection{Post-processor}
\yeounoh{
This module functions as a deterministic layer to enforce data semantic and relational constraints on the raw data generated by the LLM. It specifically targets two areas:
\\
\noindent\textbf{
Predicate and Data Constraint Enforcement:} The module applies user-defined signals and extracted predicate constraints (e.g., specific date ranges or categorical filters). Since these constraints typically affect only a small subset of the schema, the post-processor performs direct value substitution or correction on the generated instances.
\\
\noindent\textbf{
Relational Join Integrity:} Using the associations (join constraints) extracted by the pre-processor, this module synchronizes join-key values across tables. Values from a primary table are matched and copied to the corresponding columns in joined tables, ensuring that the generated mock data remains join-consistent even when the LLM's initial output lacks relational coherence.
}
\yeounoh{
By decoupling these deterministic operations from the probabilistic generation of the LLM, we significantly reduce total latency—as these updates do not require additional LLM calls -- and provide a robust mechanism to correct structural or semantic inconsistencies.
It is also important to note that enforcing join constraints during post-processing allows us to decouple the generation process at the table or schema level. By treating each table as an independent generation target, we can maximize horizontal parallelization across the workload.}

\subsection{Data validator}\label{sec:data_validator}

\begin{figure}[!th]
\centering
\caption{Semantic validation prompt template with generated test data and various extracted constraints as input.}
\label{fig:semantic_template}
\begin{tikzpicture}
    \node[draw=black!60, fill=lightgray!5, rectangle, rounded corners, inner sep=6pt, text width=0.92\columnwidth] {
            \scriptsize 
            \textbf{You are a GoogleSQL expert.} GoogleSQL is a Google internal implementation of the SQL query language.
            Given data in json format and constrains for particular columns in data, validate the data.
            \\[0.5em]

            \textbf{Instructions for validation:}
            \begin{itemize}
                \item Check the values required for a column under the constraints section.
                \item Check for the same column in the data section.
                \item If the column is not present, please return NOT PRESENT.
                \item If the columns is present, please check if the data is present in the constraint for that column.
                \item If point 4 is satisfied, please print VALID.
                \item If point 4 is not satisfied, please print NOT VALID.
            \end{itemize}
            
            \vspace{0.5em}
            
            \begin{lstlisting}[
             basicstyle=\color{promptred}\ttfamily\scriptsize,
             columns=fullflexible,
             breaklines=true,
             breakatwhitespace=true,
             frame=none,
             language=tex,
            ]
```
user_prompt = f"""
Task:
Given the following Data and Constraints, validate the data and return one of VALID, NOT VALID, NOT PRESENT.

Data:
{data}

Constraints:
{constraints}
"""
```
            \end{lstlisting}
    };
\end{tikzpicture}
\end{figure}

\yeounoh{Our framework employs a dual-stage validation pipeline to ensure both structural integrity and semantic coherence.
We validate the syntax of the generated data instances using a static parser, like \textit{proto-parser}~\cite{protobuf}. This deterministic parser ensures that the generated data adheres to the schema and satisfy all the type and structural constraints.}

\yeounoh{
Because conventional static parsers cannot capture complex business logic or test SQL-specific predicates for semantic coherence (to the test question and to other column values), we also utilize an \textit{LLM-as-a-judge}~\cite{zheng2023judging} approach. The validator receives the generated data alongside the semantic constraints extracted by the pre-processor (e.g., data ranges or current types). 
To ground the model's reasoning and resolve inconsistencies, we implement a retry mechanism. If a generation fails at validation stage, the system retries up to three times, feeding the specific error message or semantic failure reasons back into the generator to guide the next attempt. This iterative grounding reduces semantic error rates (Section~\ref{sec:semantic_validation_llm_judge}).
The template in Figure~\ref{fig:semantic_template} illustrates the detailed instructions for the judge.}

\begin{table*}[h!]
\caption{Enforced data validation rules. The generator extracts most of the syntactic and semantic constraints automatically, including the join constraints, column correlation and predicate value constraints.}
\label{tab:validation_rules}
\begin{tabularx}{\textwidth}{l >{\hsize=0.6\hsize}X >{\hsize=1.4\hsize}X}
\toprule
\textbf{Rule} & \textbf{Validation Requirements}  & \textbf{Validation Approach} \\
\midrule
$r_1$ & No missing columns & Use a proto parser to check for missing columns or proto fields. \\
\addlinespace 
$r_2$ & Valid data types &  Use a proto parser to check for invalid data types. A schema column or proto field is invalid if its data type is different from the one specified in the fetched schema/proto. \\
\addlinespace
$r_3$ & Grouped column correlation & Use statistical tools, like Pearson Correlation (numerical vs. numerical) and Chi-Square Test (categorical vs. categorical), to assert non-zero column correlations. \\
\addlinespace
$r_4$ & Join constraints & Matches any join constraints extracted in the pre-processing. The column values in the joined tables at the join columns would be matched/copied from one table data to another. \\
\addlinespace
$r_5$ & Semantic constraints & Semantic constraints are extracted from the test SQL query  and they are validated using LLM-as-a-judge.  \\
\bottomrule
\end{tabularx}
\end{table*}

Table~\ref{tab:validation_rules} summarizes various syntactic and semantic data validation rules enforced by the generator; the generator retries the generation up to three times along with the error message upon any validation failure, including the semantic ones.
In section~\ref{sec:completeness_eval}, we evaluate how often the LLM hallucinates and returns semantically invalid data that passes the validation checks (Table~\ref{tab:test_hallucnation}).

\section{Evaluation results}
We evaluate our LLM-based test data generator by investigating three key aspects that demonstrate its advantage in creating high-quality, comprehensive test datasets:
\begin{itemize}
    \item Complex Query Support: We assess the test coverage achieved, focusing on the data's utility for testing complex SQL queries.
    \item Schema Adherence: We quantitatively evaluate the data's structural integrity and fidelity to the complex nested schemas often encountered in our SQL code generation workloads, using criteria defined in Table~\ref{tab:customer_table_data_generation}.
    \item Practical Effectiveness: We compare our generated data against limited authorized production samples to demonstrate its effectiveness in covering all semantically relevant edge cases for target test SQL queries/functions.
    \item \yeounoh{Schema Format Versatility: We evaluate the methodology's robustness by applying it to JSON, a format with deeply nested structures comparable to GoogleSQL's nested types, confirming the approach is not limited to GoogleSQL and Protocol Buffer definitions}
\end{itemize}

\subsection{Evaluation setup}
Test cases are submitted by the developers who work on the code generation services. In SQL Code Assistant, a developer may generate SQL functions using the assistant service for testing GoogleSQL modules and use the generated SQL functions with some test fixtures (e.g., manually curated schema/proto instances -- often incomplete and neglect nested fields). We sampled 20 SQL Code Assistant generated test cases that require querying over complex data structures -- out of 129 disabled (buildable, but unexecutable) test cases. In NL2SQL, each submitted test case contains a natural language question and a golden GoogleSQL query, verified by engineers expert in GoogleSQL. We sampled 148 test cases out of which, only 45 had authorized access to production tables for testing.

\yeounoh{While we provide a preliminary comparison in Table~\ref{tab:customer_table_data_generation}, our primary evaluation focuses on the effectiveness of our proposed pipeline. We found that ``Manual" curation is practically impossible for large-scale industrial testing due to the time required (over an hour per test data instance), and ``Vanilla LLM" fails to generate syntactically valid data for deeply nested structures.}
\yeounoh{Furthermore, existing statistical and LLM-based tabular generators~\cite{xu2024llms,borisov2022language} were excluded as they require production data samples—which are strictly restricted in our environment—to learn the data distributions. Therefore, our method serves as a necessary zero-shot solution where prior tools fail to function.}

For our test data generation pipeline, we used  \textit{Gemini 2.5-Flash} and with 10 concurrent threads handling nested structures (e.g., nested field that leads to another schema/proto structure). The threading happens at the same level of nesting to ensure that there is no data dependency between concurrent threads.

\subsection{Test coverage with more complex SQL query testing}\label{sec:test_coverage_eval}
We assess the test coverage achieved by the generated test data. Our generation method can generate test data for complex schema with deeply nested structures. This facilitates more comprehensive testing, enabling more complex SQL test cases that were disabled due to the lack of such complex test data.


Table~\ref{tab:test_coverage_eval} presents improvement (\%) in test coverage and test accuracy -- we report the improvement instead of the raw test coverage values. 
The improvement in NL2SQL testing coverage is more significant at $69.59\%$ because many available test cases were disabled due to restricted production data access. 
Notice that the generated test data also improves NL2SQL test accuracy. To our surprise, there were many cases where manually curated production test data that failed to differentiate predicted and golden SQL queries. For instance, some aggregate queries with wrong predicted filter predicates resulted in the correct answers (false positives); many test queries with INNER JOIN were flagged as incorrect (false negatives) against the manually curated test data, whereas adding sufficient test data for all the edge cases for the relevant columns used in the SQL query (Section~\ref{sec:business_analytics_discussion}) resulted in correct results.
In SQL Code Assistant there are a lot more disabled test cases due to lack of test data-- it is challenging to generate complex nested test data instances manually. We expect to roll out new generated test data for those in the near future, which should increase the coverage further.

Table~\ref{tab:ablation_coverage} presents an ablation study for the NL2SQL pipeline, quantifying the relative impact of each component on test coverage.
Removing the Pre-processor or Schema Fetching results in a 100\% loss in coverage, as these modules provide the foundational metadata required for any data generation.
The removal of Schema Annotations leads to a 47.76\% decrease in coverage, highlighting the importance of user comments and documentation-derived context for complex, real-world schemas.
While Column Grouping and Data Validation show a 0\% impact on raw test coverage, \yeounoh{these components are critical for maintaining the semantic validity and logic required for representative evaluation results; without them, test executability remains unchanged, but data integrity is compromised.}
Removing the Post-processor reduces coverage by 20.15\%, demonstrating the necessity of deterministic join enforcement for queries involving multi-table dependencies.

We omit SQL Code Assistant from this ablation because its test cases primarily involve simple selection and projection queries; for these scenarios, basic SQL parsing and schema fetching are sufficient to achieve full coverage improvement.
\yeounoh{It is important to note that this is still infeasible with manual curation or existing static data generation tools. The LLM-based approach is essential because traditional tools cannot reason through the extreme structural complexity and deep nesting inherent in our production schemas.}

\begin{table}[t!]
\centering
\caption{Test coverage and accuracy improvements (\%) with our mock test data generation. }
\label{tab:test_coverage_eval} 
\resizebox{0.85\columnwidth}{!}{
\begin{tabular}{l c c}
\toprule
\textbf{Metric} & \textbf{SQL Code Assistant} & \textbf{NL2SQL} \\
\midrule
Coverage (\%) & +12.58 & +69.59 \\
Accuracy (\%) &+0.0 & +20.4 \\
\bottomrule
\end{tabular}
}
\end{table}

\begin{table}[!t]
\centering
\caption{Relative test coverage difference without each component/technique for test data generation in NL2SQL. }
\label{tab:ablation_coverage}
\resizebox{0.85\columnwidth}{!}{
\begin{tabular}{lcc}
\toprule
\textbf{Component Removed} & \textbf{Coverage Difference (\%)} \\
\midrule
Pre-Processor & $-100.00$ \\
Schema Fetching & $-100.00$ \\
Schema Annotation & $-47.76$ \\
Column Grouping & $0.00$ \\
Post-Processor & $-20.15$ \\
Data Validator & $0.00$ \\
\bottomrule
\end{tabular}}
\end{table}

\subsection{Generation quality evaluation with complex nested data structures}
\label{sec:completeness_eval}

\begin{figure}[!t]
\centering 
\includegraphics[width=\columnwidth]{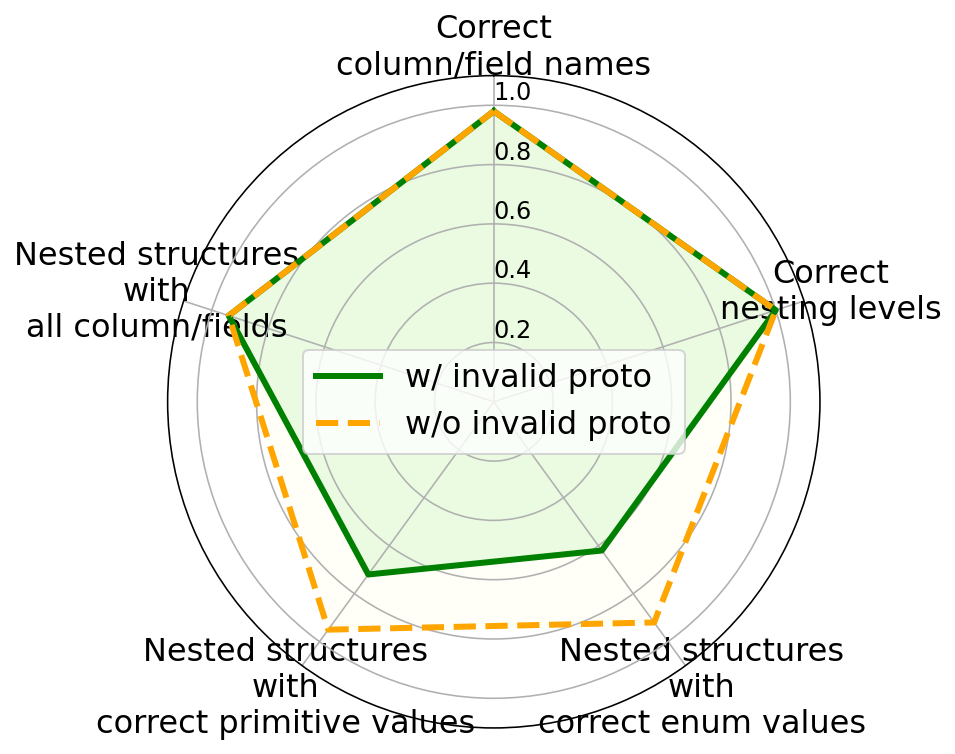} 
\caption{Structural integrity \& adherence to complex schema analysis, using key criteria from Table \ref{tab:customer_table_data_generation}. Our method achieves near perfect structual integrity if we exclude invalid schemas.}
\label{fig:structure_eval}
\end{figure}

Next, we evaluate the completeness and correctness of the generation for complex nested data formats.  Specifically, we examine the generated test data quality in terms of its structural integrity to the complex schema with columns  with deeply nested field types. 
Figure \ref{fig:structure_eval} shows the normalized results (0 to 1, higher is better), using key criteria from Table \ref{tab:customer_table_data_generation} -- we excluded the criteria for primary key generation, which is always true for our test data. Despite issues with some inaccessible (or stale) schemas in production environment, our data generation successfully produces test data compliant with complex, real-world nested structures.

Overall, our generated test data adheres very well to complex schema with deeply nested column types encountered in production SQL code generation services. There are some cases where nested field (proto) definitions are stale or removed, in which case the LLM guesses the field names and types for the inaccessible nested proto structures. Note that the  scores are much higher (and near perfect) without such invalid proto definition cases \textit{w/o invalid proto}. Note that the \textit{correct nesting levels} scores for both with and without invalid proto are the same because we skip over those invalid nested structures in the aggregation -- since we cannot assess the correct nesting levels for unknown proto. We typically see 3-level of nesting in complex SQL testing scenarios, and as deep as 5 levels.

\subsection{Semantic data constraints validation with LLM-as-a-judge}
\label{sec:semantic_validation_llm_judge}

\yeounoh{The generator extracts semantic data constraints, like which column values are relevant to the given SQL query in the pre-processing step. 
We use the LLM as LLM-as-a-judge to validate the generated test data against the extracted semantic constraints. If some generated data instances fail to satisfy the constraints according to the LLM, then we retry with an additional hint that the previous generation failed to satisfy the failed semantic constraints. In all test cases, we have successfully generated test data instances after 1 or 2 retries. Interestingly, we observed that the LLM can hallucinate and actually pass invalid data instances as valid. 
Table~\ref{tab:test_hallucnation} illustrates how generating multiple data instances per test case can limit the hallucination error rates, defined as the proportion of test cases with no semantically valid test data instances after $n$ instances generated.
We report the hallucination error rates for NL2SQL test cases only because the unit test cases for SQL Code Assistant are too simple and contains simple semantic constraints, like COUNTIF(experiment\_group = ``GROUP A") AS group\_a\_count. 
It is important to note that without schema-level annotations (e.g., column grouping), the pipeline often generates semantically irrelevant test data instances (more than 88\% of the time). Interestingly, the few successful test cases were simple aggregation or select queries that required only a handful of syntactically valid rows in the tables.
While most semantic constraint violations are resolved with a single additional data generation attempt ($n=2$), a few constraints remain challenging for the LLM-as-a-judge to validate. Specifically, complex datetime filter constraints that require type casting, time zone conversion, and date extraction consistently failed even after five attempts ($n=5$). For example, consider the constraint  DATE(TIMESTAMP\_SECONDS(CAST(status.time\_processed
\_sec AS INT64)), 'America/Los\_Angeles') $\geq$ ``2023-07-01". 
Given a generated value like "1678886400" for time\_processed\_sec, the validator correctly performed the initial conversion, recognizing that the timestamp corresponds to the date ``2023-03-15". 
However, it then failed to perform the final step of evaluating whether this date satisfied the constraint (i.e., that ``2023-03-15" is not greater than or equal to ``2023-07-01"). A potential remedy involves invoking a reasoning model or chain-of-thought prompting~\cite{wei2022chain}, though this significantly increases generation time. We plan to explore more cost-efficient ways to address these hallucination errors in future work. 
}

\begin{table}[t!]
\centering
\caption{LLM hallucination error rates for semantic data validation with repeated data generation for each test. By generating multiple instances per test $n>1$, we get at least one test instance that satisfies the semantic constraints. The results for SQL Code Assistant was perfect, since the unit test cases does not involve any complex semantic constraints. }
\label{tab:test_hallucnation}
\resizebox{0.85\columnwidth}{!}{
\begin{tabular}{l c c c c c}
\toprule
& \multicolumn{5}{c}{\textbf{Number of instances per test ($n$)}} \\
\cmidrule(lr){2-6}
& 1 & 2 & 3 & 4 & 5 \\
\midrule
NL2SQL & 6.08\% & 2.02\% & 2.02\% & 2.02\% & 2.02\% \\
\bottomrule
\end{tabular}
}
\end{table}

\subsection{A comparative study VS. production sample}
\label{sec:diversity_eval}
Here we evaluate the effectiveness of our data generation methodology in comparison to production data sampling. We use one  production table authorized for testing to generate data samples of different sizes. To compare the quality of our generated test data against the production samples, we compare the number of distinct values of the relevant predicates/columns extracted from the question and golden SQL query. We use a test case submitted by an anonymous developer, ``How many Shorts videos about Gaming were uploaded by (identified as) female creators in the United Kingdom(GB) during Q1 2024?," for the test production table. 
There are two relevant columns also used as predicates, describing different country codes and semantic tags for the Shorts videos. 

\begin{figure*}[t!]
    \centering
    \begin{subfigure}{\columnwidth} 
        \centering
        \centering
        \includegraphics[width=0.9\columnwidth]{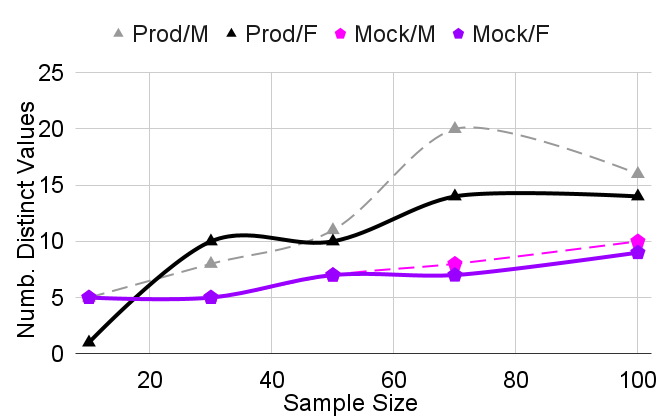} 
        \caption{Country codes}
        \label{fig:subfig1}
    \end{subfigure}
    \hfill 
    \begin{subfigure}{\columnwidth} 
        \centering
        \includegraphics[width=0.9\columnwidth]{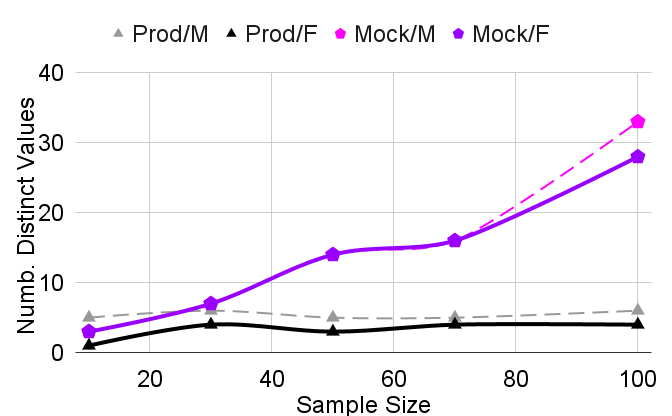} 
        \caption{Semantic tags}
        \label{fig:subfig2}
    \end{subfigure}
    \caption{The number of distinct values generated for relevant predicates columns in the test SQL query and the natural language question, "How many Shorts videos about Gaming were uploaded by (identified as) female creators in the United Kingdom (GB) during Q1 2024?" }
    \label{fig:sample_study}
\end{figure*}

Figure~\ref{fig:sample_study} presents the number of distinct values for the (a) country codes column and (b) semantic tags column. Because the question is asking for the (identified as) female case, we also track the counts by (identified as) male and (identified as) female. In case of country codes, the production samples \textit{Prod} include a more diverse set of countries than the \textit{Mock} ones. This is OK because the question is looking for the answer in the United Kingdom (GB) specifically -- and it is likely that our generation is less diverse because of this. At the same time, we also notice an imbalanced sampling between \textit{Prod/M} and \textit{Prod/F} in the production samples, which is not desirable as the question seeks to answer for the (identified as) female case.
In case of semantic tags, which can entail different genres and keywords for different gaming contents (FPS, MMO, RPG, tutorial, review, etc.), our generated test data \textit{Mock} does include a more diverse set of values. In both cases, the data is well balanced between the two select genders. Note that the test data is generated with syntax and semantic validations as described in Section~\ref{sec:data_validator}, but the values can still be irrelevant as it is not learned from the actual production data (not available). Our generation approach also takes the test query/function into account to generate data that is more semantically aligned to the query. We discuss another example in Section~\ref{sec:business_analytics_discussion}.

\subsection{\yeounoh{Generalizability to Other Data Formats}}

\begin{table}[!th]
\centering
\caption{\yeounoh{Comparison of our method, using \textit{Gemini 2.5-Flash} against SOTA models on the DeepJSONEval benchmark (Overall metrics) leaderboard.}}
\label{tab:deepjsoneval}
\begin{tabular}{@{}lccc@{}}
\toprule
\textbf{Model} & \textbf{Format} & \textbf{Detailed} & \textbf{Strict} \\ \midrule
\textbf{Our Method (Gemini 2.5-Flash)} & \textbf{100.0} & \textbf{92.11} & \textbf{59.86} \\
Claude Sonnet 4 & 99.05 & 90.73 & 57.90 \\
Magistral Medium 2506 & 98.10 & 90.36 & 59.81 \\
DeepSeek R1 0528 & 97.90 & 89.57 & 57.33 \\
Gemini 2.5 Pro & 97.52 & 89.00 & 56.19 \\
Qwen3 235B A22B & 97.14 & 88.33 & 56.19 \\ \bottomrule
\end{tabular}
\end{table}

\yeounoh{
To demonstrate the generalizability of our approach beyond GoogleSQL and proto, we evaluated our pipeline using the \textit{DeepJSONEval} benchmark~\cite{zhou2025deepjsoneval}. This dataset focuses on generating deeply nested JSON objects that must strictly adhere to complex schemas and semantic constraints.
}
\yeounoh{As shown in Table~\ref{tab:deepjsoneval}, our method leveraging \textit{Gemini 2.5-Flash} achieves the best results, even outperforming those by more advanced SOTA models. This result confirms that our pre- and post-processing modules effectively handle diverse nested structured data formats without dialect-specific tuning.}

\section{Related works}

\noindent\textbf{Test data generation for complex SQL testing:} Test data generation is crucial for SQL query validation, query rewriting, automated query grading, and NL2SQL tasks. Common strategies involve using manually curated test data, production data sampling, constraints based data generation for table schema and SQL query~\cite{somwase2024data,whiting2008creating,bhangdiya2015xda,chandra2015data,arcuri2019restful}. 
Recent research~\cite{singh2024exploring} has explored the utility of LLMs for generating artifacts critical to database testing and development, like diverse and complex SQL test queries. 
This area, along with the broader application of LLMs in test case generation~\cite{chen2024chatunitest,schafer2023empirical}, is attracting significant interest.
This line of research, and the related benchmarks, is primarily concerned with leveraging LLMs to generate a diverse test SQL queries beyond a handful of simple SQL equivalent pairs; however, this is currently done using rather simple, non-production database schema.

In contrast, our work addresses a distinct but equally critical challenge: the complex and relevant  generation of the test data for large complex, nested production schema structures -- with relevant data instances that adhere to intricate real-world constraints. This is essential for effectively unit-testing complex SQL functions and evaluating the robustness of NL2SQL systems at Google.
The key challenges in generating high-fidelity test data is handling more complex nested schema structures and their inter-dependencies. Many SQL equivalence testing and constraints based SQL test data generation do not support arbitrarily nested structures and nested fields (array, struct and/or proto types) \cite{he2024verieql,somwase2024data} -- limiting the test use cases to less complex SQL queries.
Production data sampling can provide the most realistic high-fidelity test data for any complex schema, but it is often prohibited or infeasible in industrial contexts where service providers lack access to user data due to stringent privacy and security protections/policies. 
\\
\noindent\textbf{Model-based tabular data generation:} 
Despite emerging interest in using LLMs for test data generation \cite{gretel_definitiveguide,lu2023machine,karmarkar2024navigating,liu2023fill}, existing approaches often fail to meet the complex requirements of industrial testing. While LLMs can handle basic schema adherence~\cite{borisov2022language,narayan2022can,korini2023column,peeters2023schema,hu2024efficient,li2024automatic}, out-of-the-box models struggle with subtle data relationships \cite{xu2024llms} and become incomplete for large, nested schemas (Table \ref{tab:customer_table_data_generation}).
\yeounoh{One of the major barriers to using existing LLM-based tabular generators in our setting is their reliance on actual table samples to learn data distributions (e.g., fine-tuning). Our approach addresses this gap by operating in a true cold-start regime. Instead of requiring seed data, we leverage the LLM’s internal reasoning by using the natural language question and the SQL query body as primary signals to generate semantically relevant data.}
Similarly, traditional deep generative models \cite{zhang2023mixed,lee2023codi,liu2023goggle} focus on flat schemas and require extensive training data, which is often restricted in real-world environments.
To address these limitations, we propose a complex test data generation method based on the pre-trained LLM, \textit{Gemini}, requiring the testing function/SQL statements as well as the corresponding business logic questions (NL2SQL).

\section{Discussion}

\subsection{Scaling and cost of the generation}
\label{sec:cost_analysis}

\begin{table}[t!]
\centering
\caption{Average accumulated tokens and generation time per instance for \textit{Gemini 2.5-Flash}.}
\label{tab:cost_analysis}
\resizebox{\columnwidth}{!}{
\begin{tabular}{l c c}
\toprule
\textbf{Metric} & \textbf{SQL Code Assistant} & \textbf{NL2SQL} \\
\midrule
Input tokens & 162,199.5 ($\pm$ 350,852.1) & 57,082.3 ($\pm$ 102,381.8) \\
Output tokens & 3,890.0 ($\pm$ 3,121.5) & 8,054.2 ($\pm$ 13,040.1) \\
Time (sec) & 66.0 ($\pm$ 53.6) & 57.5 ($\pm$ 64.36) \\
\bottomrule
\end{tabular}
}
\end{table}

\begin{table}[htbp]
\centering
\caption{\yeounoh{Data diversity across temperatures for a 10,000-row generation.}}
\label{tab:diversity_temperature}
\resizebox{\columnwidth}{!}{
\begin{tabular}{lcccc}
\toprule
\textbf{Metric} & \multicolumn{4}{c}{\textbf{Temperature}} \\ \cmidrule(l){2-5} 
 & \textbf{0.1} & \textbf{0.5} & \textbf{1.0} & \textbf{1.8} \\ \midrule
Unique instances at 10k & 10,000 & 10,000 & 10,000 & 10,000 \\ \addlinespace
\textbf{video\_sid uniqueness} & 50\% & 50\% & 40\% & 70\% \\ 
\textbf{country\_code coverage} & 87.14\% & 7.2\% & 39.75\% & 44.17\% \\ 
\textbf{Enum field coverage} & 100\% & 100\% & 100\% & 100\% \\ \bottomrule
\end{tabular}
}
\end{table}


\yeounoh{
Our automated pipeline is highly parallelizable across query batches and deeply nested structures, addressing the primary industrial bottleneck where manual data curation time often exceeds test execution time (Table~\ref{tab:customer_table_data_generation}). Table~\ref{tab:cost_analysis} quantifies the efficiency leap: while a domain expert requires over an hour to curate a single comprehensive instance for complex schemas, our framework generates high-fidelity data in 57.5--66.0 seconds on average. The high variability in input tokens reflects the diverse schema complexities, including recursive nested  structures, successfully handled by the system without production data access.
}

\yeounoh{
To assess scalability for large-scale workloads, we evaluated the system's ability to maintain diversity across 10,000 generated instances (Table~\ref{tab:diversity_temperature}). The pipeline sustains high quality without significant repetition as dataset size increases. For high-cardinality fields like \textit{video\_sid}, the model achieved up to 70\% uniqueness. While most SQL validation tasks require only a few hundred unique identifiers, our framework supports deterministic post-processing—such as ID tracking or incremental assignment—to ensure absolute uniqueness for datasets exceeding 10,000 rows. For constrained fields like country codes, the LLM explored up to 87.14\% of the value space, while enum coverage remained perfect (100\%) across all temperatures. These results confirm that the method produces semantically grounded, diverse datasets suitable for industrial-scale SQL testing.
}

\subsection{Generating for test SQL query semantics}
\label{sec:business_analytics_discussion}


\begin{figure}[!htbp]
\centering
\begin{tikzpicture}
  \node[draw=black!60, fill=lightgray!5, rectangle, rounded corners, inner sep=6pt, text width=0.92\columnwidth] {
            \begin{lstlisting}[basicstyle=\ttfamily\scriptsize, columns=flexible, showstringspaces=false]
SELECT date_trunc(logdate, quarter), sum(conversions) / sum(clicks) 
FROM fake_table_1
WHERE result_type = 'TEXT_AD' AND logdate BETWEEN '2022-01-01' AND '2022-12-31' 
GROUP BY date_trunc(logdate, quarter);
            \end{lstlisting}
        };
  
\end{tikzpicture}
\caption{High-fidelity test data generation for this business analytics query requires generating data instances for each quarter in year 2022.}
\label{fig:quarterly_report_query}
\end{figure}

Semantic coherence between test data and the target SQL query is essential for evaluating SQL code generation services like NL2SQL. High-fidelity test data must move beyond structural integrity to reflect the functional logic of the query. For instance, the business analytics query in Figure~\ref{fig:quarterly_report_query}, which calculates quarterly conversion rates, requires data instances distributed across every quarter of 2022 to produce a valid report.
Our LLM-based generator can recognize such semantic importance of \textit{date\_trunc(logdate, quater)} and generate data for all quarters in the range. Moreover, if one has a domain knowledge about the data, such as ``conversions were better in Q4 than Q1," then it can be passed to the generator as \textit{user\_criteria} as part of input, which can guide the generation to favor Q4 instances.
Such understanding and reasoning capabilities of LLMs can scale semantic-aware test data generation.

\subsection{Generation error analysis}
\label{sec:error_analysis}
Our evaluation revealed several recurring error patterns in the generation process: formatting inconsistencies (invalid JSON), structural omissions (missing fields or rows), and hallucinations where extraneous columns were introduced—particularly in simpler schemas. Most of these issues are effectively mitigated through automated retries and schema-tracking safeguards that enforce strict adherence to the requested field list.

Addressing missing edge cases for complex SQL predicates remains a challenge without production data or an oracle. While our \textit{LLM-as-a-judge} validator provides semantic verification, it is susceptible to false positives. To improve fidelity, we recommend integrating domain-specific constraints (as discussed in Section \ref{sec:business_analytics_discussion}) to guide the LLM toward relevant edge cases. Future investigation into statistical imputation and estimation techniques \cite{biessmann2019datawig,chung2018estimating} could further enhance mock data quality in production environments.

\section{conclusion and future work}
In this paper, we addressed the critical challenge of generating semantically relevant and complex test data for Google SQL code generation services, particularly where production data access is restricted. We highlighted the limitations of traditional data generation methods that were used by developers, focusing on their inability to handle large, deeply nested data structures, like protocol buffers, and to maintain semantic integrity of the target test cases. 
By leveraging the \textit{Gemini 2.5-Flash} model with strategic pre- and post-processing, we successfully generated comprehensive test data, compliant with large, complex data structures. This data effectively supports complex SQL testing, including queries with joins, aggregations, and nested subqueries, and significantly improves test coverage.
Our experiments demonstrate a reliable and practical method for generating high-fidelity test data, which is essential for industrial environments where access to production data is restricted. While our current implementation targets Google SQL, the underlying methodology is dialect-agnostic.


\section*{Acknowledgment}
We would like to thank John Cieslewicz, Kashmira Phalak, Yannis Papakonstantinou and the anonymous reviewers
for their valuable feedback. We also thank Yuchen Zhang for his contribution in the early stages of this project.


\bibliographystyle{IEEEtran}
\bibliography{main}

\end{document}